# The Requirements for Ontologies in Medical Data Integration: A Case Study

Ashiq Anjum, Peter Bloodsworth, Andrew Branson, Tamás Hauer, Richard McClatchey, Kamran Munir, Dmitry Rogulin, Jetendr Shamdasani

*CCS Research Centre, CEMS Faculty, University of the West of England, Coldharbour Lane, Frenchay, Bristol BS16 1QY, UK*
*Email: {Ashiq.Anjum, Peter.Bloodsworth, Andrew.Branson, Tamas.Hauer, Richard.McClatchey, Kamran.Munir, Dmitry.Rogulin, Jetendr Shamdasani}@cern.ch*

**Abstract**

*Evidence-based medicine is critically dependent on three sources of information: a medical knowledge base, the patient's medical record and knowledge of available resources, including, where appropriate, clinical protocols. Patient data is often scattered in a variety of databases and may, in a distributed model, be held across several disparate repositories. Consequently addressing the needs of an evidence-based medicine community presents issues of biomedical data integration, clinical interpretation and knowledge management. This paper outlines how the Health-e-Child project has approached the challenge of requirements specification for (bio-) medical data integration, from the level of cellular data, through disease to that of patient and population. The approach is illuminated through the requirements elicitation and analysis of Juvenile Idiopathic Arthritis (JIA), one of three diseases being studied in the EC-funded Health-e-Child project.*

**Keywords:** Biomedical application, user requirements, data modelling, ontologies

# 1. Introduction

## 1.1. The problem in general

Information technology today is widely adopted in modern medical practice, especially supporting digitized equipment, administrative tasks, and data management but less has been achieved in the use of computational techniques to exploit the medical information in research or practice. There is an emerging demand for the integration and exploitation of heterogeneous biomedical information for improved clinical practice, medical research and personalized healthcare [1].

The application of computational and mathematical techniques in applying the laws of medicine to solve clinical problems is difficult because these laws are not mathematical in nature. Biology and especially medicine are "knowledge based disciplines" relying greatly on observed similarities rather than on the application of precise rules [2], [3]. Even though the amount of digitized – and thus computer accessible – information is substantial and increasing, effective use of computer-aided processing of the data is usually not feasible because of the vast amount of implicit semantics behind that information. The lack of explicit semantics is also responsible for heterogeneity which makes integration particularly difficult.

The Health-e-Child (HeC) project aims to demonstrate that indeed integrating medical integration in novel ways yields immediate benefit for clinical research and practice. We are studying the use of integration along different integration axes including the following examples. The first aspect is that of sharing data among spatially separated clinicians – bringing together information produced in different departments or multiple hospitals for the purpose of creating statistically significant samples, studying population characteristics and sharing knowledge among clinicians. Vertical integration means establishing a coherent view of the child's health to which information from each vertical level contributes, from molecular through cellular to individual. Temporal integration is particularly interesting. The conventional patient record is a bag of "snapshot" information. Disease progression, deterioration and change quantification are of the utmost importance, especially with chronic diseases. Supporting the time axis is particularly challenging in paediatrics as change quantification must account for the normal growth of the patient, too. The really interesting questions come from the combination of these axes, for example "can one predict from a genetic profile the expected joint deterioration over five years in certain populations". We have recently studied how these integration challenges reflect in concrete requirements in the context of the Health-e-Child project. In this paper we give an account of some of those related to paediatric rheumatology, in particular we present the concrete example of Juvenile Idiopathic Arthritis.

In the next sections we introduce the Health-e-Child project and detail its high level user requirements. Then in Section 2 the JIA case study is introduced and its specific requirements are investigated. Section 3 describes the approach being adopted in HeC to address the low-level JIA requirements and we indicate in the conclusions how this approach can be used in the other clinical areas being studied in HeC.

### 1.2. The Health-e-Child project

There is a compelling demand for the integration and exploitation of heterogeneous biomedical information for improved clinical practice, medical research and personalised healthcare across the EU. The Health-e-Child project [1] aims to develop an integrated platform for European Paediatrics, providing seamless integration of traditional and emerging sources of biomedical information as part of a longer-term vision for large-scale information-based research and training, and informed policy making. The HeC project will study the data integration requirements of clinicians treating paediatric heart diseases, Juvenile Idiopathic Arthritis (JIA) and child brain tumours and will provide demonstrators for decision-support, knowledge management and disease modelling. Clinicians from the hospitals of the Gaslini Institute (Genoa, Italy), Great Ormond Street (London, UK) and the Necker Institute (Paris, France) are working with computer scientists, knowledge engineers and bioinformaticians (both academic and commercial) to produce project deliverables in a 4-year Integrated Project initiated at the start of 2006, supported by the eHealth Unit of the EC's ICT Framework 6 programme. The first demonstrable prototypes demonstrating data integration using a Grid platform are scheduled for review and feedback from the clinical community towards the end of 2007.

### 1.3. The High-level requirements of HeC

In order to ensure agreement on realisable project objectives, to establish the foundations for clinical collaboration in the prototype development process and to provide understanding of the clinical processes and constraints inherent in the HeC study, a process of rigorous requirements elicitation was initiated at the project outset. This involved a number of elicitation meetings comprising study of existing practice, interviews with clinicians, observation of current system usage and evaluation of potential technological options. Use-case and conceptual data models were incrementally and iteratively developed and validated as the main requirements models in consultation with key clinical staff in order to ensure that HeC requirements engineering process addressed the needs of the specific communities of paediatric heart diseases, JIA and child brain tumours.

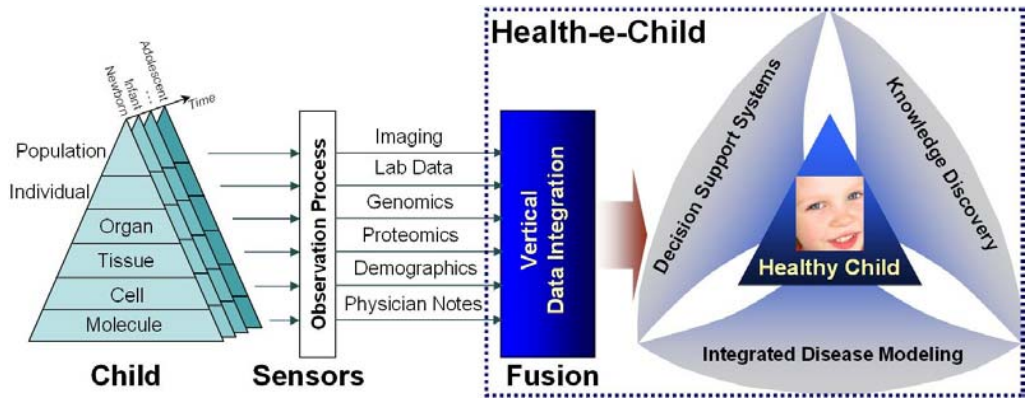

Figure 1: The conceptual approach of 'vertical data integration' in Health-e-Child

Health-e-Child information sources are spread over a 'vertical' dimension across disciplines (biomedical abstractions from genetics to epidemiology, see figure 1), a set of lateral dimensions between physical data sources, a temporal dimension and a conceptual abstraction (from data to information to knowledge). Data is linked through the individual patient it relates to, providing the basis for personalised treatment; collecting the many different types of information that is available about an individual into a coherent whole.

A number of sophisticated analysis and modelling techniques are required to address the specialised needs of Health-e-Child applications synthesising disease models, knowledge discovery and data mining algorithms, biomedical similarity measures and ontology integration mappings. Health-e-Child will adopt a conservative attitude towards technology, developing only what is strictly necessary. This is best achieved through a system developed on top of generic infrastructures that provide all the required common data and computation management services. The emphasis of the Health-e-Child requirements process is therefore on "universality of information" and its corner stone is the integration of information across biomedical abstractions, whereby all layers of biomedical information are 'vertically integrated' (see figure 1) to provide a unified view of a child's biomedical and clinical condition. The next section introduces JIA – one of the integration areas studied in HeC and one that illustrates the importance of 'vertical' integration particularly well.

## 2. Paediatric Rheumatology

### 2.1. Juvenile Idiopathic Arthritis

Juvenile Idiopathic Arthritis (JIA) [4] has a resemblance to Rheumatoid Arthritis, but with an onset in childhood (defined as before 16 years of age). The disease is assumed to have an auto-immune mechanism and is characterized by a generalized inflammation affecting multiple joints, but there is a wide range of conditions that are collectively labelled JIA and the group of patients is not uniform. Although the disease is thought to have molecular or cellular origin, the accepted diagnosis of JIA is mostly clinical.

JIA is a chronic disease with active periods that may involve a number of joints, though not necessarily the same ones every time, leaving permanent damage that may result in serious functional disability in the long term. The evaluation of the disease activity is based primarily on clinical and lifestyle assessment scores (pain, limited range of motion, swelling), while the damage progression is assessed by a doctor's evaluation of images (X-ray, ultrasound, MRI), the standardization is achieved by various scoring schemata. There is no specific therapy to cure JIA as the cause of the disease is not known.

### 2.2. Clinical requirements for modelling JIA data

It is apparent that the strategy for tackling JIA in the clinical practice relies on data spanning a wide range of 'vertical' heterogeneity: molecular, organ-specific, and lifestyle information all contribute to various aspects of diagnosis, evaluation, and treatment. This is the case with many other diseases, but in JIA the lack of sufficient understanding of the linkage of information across the vertical axis and modalities is very apparent and interesting to study. Some of the outstanding questions that we identified as candidate issues to study are:

- The diagnosis is clinical, mostly negative, while the origin is probably molecular. Assessments are based on imaging, while the treatment is local (molecular, organ), as well as global (molecular). How do these interact?
- How can the clinical variables for disease activity and damage be standardised?
- Correlations can be made between patient assessment and disease progression. Which are the early predictors of damage?
- What are the short and long term effects of treatment procedures?
- Are there markers which distinguish between reversible and permanent organ damage?

- Can we identify more homogeneous groups of patients for better classification of JIA subtypes and better planning for treatment response?
- Are there any hereditary elements, do these correlate with other autoimmune diseases?

To answer these questions, clinicians need to test their theories on a statistically significant representation of the population. Of considerable importance are the so-called "fishing" experiments: individual centres generating ideas based on some individual case reports that can be tested against larger population or clinical trials.

## 3. Addressing JIA Requirements in Health-e-Child

### 3.1. HeC information technology solutions addressing JIA problems

The requirements mentioned in the previous sections drive the development of the Health-e-Child prototype system. There are many aspects in addressing these requirements from the end-user applications (e.g. decision support system, knowledge discovery etc.) and the data management platform points of view. For example, while the former are primarily specialized algorithm-based tools focusing on the processing of a particular (pre-defined) dataset, the latter is more concerned with serving the applications in more generic way providing the mechanisms for storage, access, management and integration of any available biomedical data and knowledge. Thus the solutions proposed by different activities within the project may differ and in effect pose new requirements on the system from an IT point of view but at the end all these serve the same purpose fulfilling the needs of the clinicians.

From the applications' point of view the following techniques have been proposed in order to address the clinical requirements with respect to JIA:

- Knowledge Discovery (KD) algorithms using statistical learning and data mining techniques to find a better classification of the JIA (homogeneous groups of patients) based on the joint analyses of all available heterogeneous biomedical data (e.g. clinical, imaging, genomics, proteomics). Currently the JIA diagnoses and treatment are primarily based on the clinical assessment. The imaging and laboratory tests assist clinicians in the evaluation of the disease progression and drug outcome but do not serve as early predictors or indicate the patient's sensitivity to a particular treatment or risk of poor outcome.
- Image-based techniques including semi-automatic segmentation of the synovitis to speed-up and improve the scoring (an inflamed synovia might be an early predictor of the severity of the disease); semi-automatic evaluation of joint damage; image registration across time to be able to compare the different studies of the same patient or between different (groups of) patients.

- Decision Support System (DSS) for individualized evaluation/treatment by monitoring disease progression and treatment outcome based on available biomedical inputs and previously established clinical knowledge.

These applications place additional requirements on the data management facilities provided by the platform. For example, the DSS system must be trained on large amounts of sample data, preferably coming from different populations; the knowledge acquired by the clinicians needs to be shared across different clinical communities; the information stored in the hospital databases should be made available for sophisticated query processing engine to assist the applications as well as to enable end-users to analyse the data.

### 3.2. Modelling requirements

In order to support the requirements discussed so far, the models for representing appropriate biomedical information need to meet specific requirements. In the context of the Health-e-Child project, these requirements are specified in a joint effort with the participating clinicians. To support the Health-e-Child objectives, a set of models for representing biomedical information need to be in place. At the first approximation, (borrowing the terminology from description logics), the information at hand may be split into terminology and assertions. In both cases, we need to observe constraints: existing knowledge should be reused, at the same time the knowledge base should be aligned to existing data models (patient record formats, examination templates, clinical protocols).

For the example at hand, various existing ontologies can contribute to representing the domain knowledge for rheumatology or JIA in particular. GALEN [5], Gene Ontology (GO) [6], FMA [7] etc. cover fragments representing population, patient semantics, or define the concepts of and relationship between joint, synovial membrane, fluid, proteomics, etc. Alignment with these ontologies is both very beneficial and challenging at the same time. On the one hand, one gains powerful reasoning support over the knowledge base and makes the acquired knowledge universal, shareable and reusable. On the other hand, the selection and alignment of the reusable fragments from different ontologies is a difficult task. In Health-e-Child a further important requirement is to capture the aspects that are specific to paediatrics: many of the existing ontologies do not directly apply to the anatomy, physiology and pathology of children.

Rather than attempting to give a complete account of the modelling requirements that we have determined, let's study three examples. The following questions have been identified as part of the set of rheumatology use-cases in Health-e-Child.

1. Identify those MRI baseline measures (degree of synovitis, bone marrow oedema etc) that are most predictive of future severe radiological damage. (Correlation of the MRI baseline variables with the change in the erosion score from baseline to 1- 2 years.)
2. Identify more homogeneous groups of patients (suitable for aetiopathogenetic studies) taking account clinical assessment, immunologic, genetic, proteomic and radiological findings.
3. Identify a panel of candidate protein biomarkers in JIA that can predict which patients will develop erosive, disabling diseases. (Are the serum and synovial protein profiles reflective of active disease and /or predictive of disease course?).

All these examples stress the importance of the integrating the diverse medical data (clinical, epidemiological, imaging, genomic, proteomic etc.) that represent the patient's information at different levels of granularity (vertical levels). The semantic link between these levels is obvious: entities are in a part-of hierarchy, but the extreme complexity of the human body and its processes usually do not allow for drawing straightforward conclusions from parts to the whole. To establish a basis for the semantic coherence of the integrated data and facilitate availability and accessibility of external information for querying and analysis by clinicians, the mappings from clinical data to the external medical knowledge (e.g. biomedical ontologies and databases) should be provided. For example, to answer the third question the clinical data needs to be aligned with the external knowledge sources to identify genetic markers that can be present in JIA.

Another important requirement for our modelling is the temporal dimension. Time is a key issue in paediatric research and practice. Clinicians are usually interested in analysing patients' data over time (see the first example). The paediatrics domain adds an additional complexity due to the fact that the child is growing and the observations in time should be aligned with the anatomical changes as well as the knowledge about how a particular disease may affect these changes. The clinical process usually follows a given time order (symptoms, study, diagnosis, treatment, follow-up, etc.). In addition, some symptom/diagnosis and treatment concepts are time-related (for example, a diagnostic criterion for JIA is persistence of some symptoms for a given time; medication is prescribed with a time profile, etc).

## 4. Ontologies in Health-e-Child

The standard textbook definition of an ontology is a formal specification of a shared conceptualisation. This means that an ontology represents a shared, agreed and detailed model of a problem domain. We are currently developing and investigating an ontological approach to represent the HeC domain. One advantage for the use of ontologies is their ability to resolve any semantic heterogeneity that is present within the data. Ontologies define links between different types of semantic knowledge. They can particularly aid in the resolution of terms for queries and other general search strategies, thus improving the search results that are presented to clinicians. The fact that ontologies are machine processable and human understandable is especially useful in this regard [8]. A complete discussion of ontologies is beyond the scope of this paper; the interested reader is referred to [9]. There are many ontologies in existence today especially in the biomedical domain, however they are often limited to one level of what we refer to as vertical integration. For example consider the Gene Ontology (GO) [6] which only defines structures regarding genes and GALEN [5] that is limited to anatomical concepts. In both cases there are no links to the other vertical levels that we have defined. We are currently investigating the scope for reusing these ontologies, or parts thereof, which have been identified by experts in both knowledge representation and clinical matters.

Many of the ontologies that exist today do not cover the paediatrics domain, to a thorough extent, for example there is a difference between the physiology of a fully grown adult and that of a child; there are also some similarities for example they both have a two lungs. Hence, it would not be sensible to reuse these ontologies in their entirety; instead we propose the extraction of the relevant parts and then the integration of these into a coherent whole, thereby capturing most of the HeC domain. However it should be noted that integrating these ontologies into one single (upper level) ontology will not be sufficient to capture the entire HeC domain, and therefore we will have to model the missing attributes and extend these existing ontologies to suit our needs. Although there are other upper level ontologies present today, such as DOLCE [10] and SUMO [11], they are considered to be too broad to be included in the project.

The ontology modelling process is known as ontology engineering. The traditional ontology engineering process is an iterative process consisting of ontology modelling and ontology validation [12]. Taking this view of ontology engineering, we have chosen to evaluate different methodologies available to us for the development of our vertical domain model. There are many methods available in the literature, for example CommonKads [13] and Diligent [14]; this evaluation process is ongoing. A methodology that deserves special consideration in this paper is proposed by Seidenberg and Rector [15] in which a strategy for modular development of ontologies is proposed, to support the re-use, maintainability and evolution of the ontology to be developed. This methodology consists of untangling the ontology into disjoint independent trees which can be recombined into an ontology using definitions and axioms to represent the relationships in an explicit fashion. To facilitate this modular methodology we are also taking under consideration the use of a fragment-based approach for the development of our domain models.

### 4.1     Ontologies and data integration

Data integration is the process of using a conceptual representation of the data and of their relationships to eliminate possible heterogeneities. Ontologies are extensively used in data integration systems because they provide an explicit and machine-understandable conceptualization of a domain. There are several approaches to data integration which we will now consider in further detail as described by Wache et al. in their article [16]. In the single ontology approach, all source schemas are directly related to a shared global ontology that provides a uniform interface to the user. However, this approach requires that all sources have nearly the same view on a domain, with the same level of granularity. A typical example of a system using this approach is SIMS [17]. In the multiple ontology approach, each data source is described by its own (local) ontology separately. Instead of using a common ontology, local ontologies are mapped to each other. For this purpose, an additional representation formalism is necessary for defining the inter-ontology mappings. The OBSERVER system [18] is an example of this approach. In the hybrid ontology approach, a combination of the two preceding approaches is used. `In the hybrid approach a local ontology is built for each source schema, which is not mapped to other local ontologies, but to a global shared ontology. New sources can be added with no need for modifying existing mappings. The layered framework [19] is an example of this approach.

The single and hybrid approaches (see figure 2) are appropriate for building central data integration systems, the former being more appropriate for so-called Global-as-View (GaV) systems and the latter for Local-as-View (LaV) systems. One drawback associated with the single global approach is the need for maintenance when new information sources are added to the representation. The hybrid architecture allows for greater flexibility in this regard with new sources being represented at the local level. The multiple ontology approach can be best used to construct pure peer-to-peer data integration systems, where there are no super-peers.

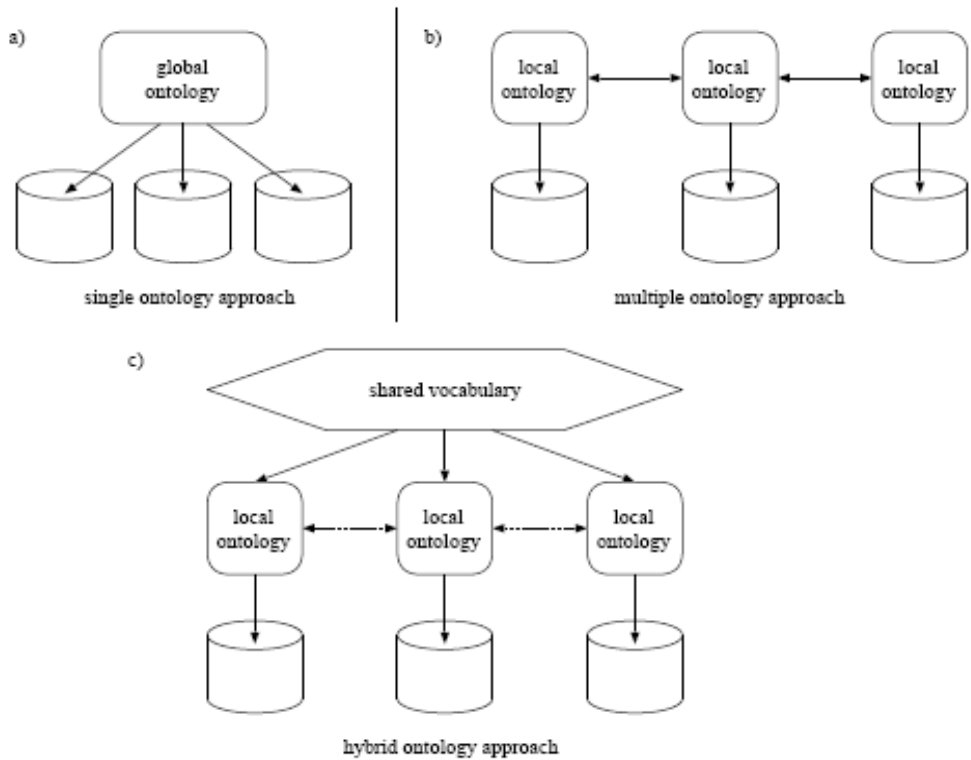

Figure 2: The three possible ways for using ontologies for data integration [16]

A mapping discovery process involves identifying similarities between ontologies in order to determine which concepts and properties represent similar notions across heterogeneous data samples in a (semi-)automatic manner. One of the major bottlenecks in generating viable integrated case data is that of mapping discovery. There exist two major approaches to mapping discovery. A top-down approach is applicable to ontologies with a well-defined goal. Ontologies usually contain a generally agreed upper-level (top) ontology by developers of different applications; these developers can extend the upper-level ontology with application-specific terms. Examples from this approach are the Suggested Upper Merged Ontology, SUMO [11] from the IEEE Standard Upper Ontology Working Group and DOLCE [10].

A heuristics approach uses lexical structural components of definitions to find correspondences with heuristics. For example, [20] describes a set of heuristics used for the semi-automatic alignment of domain ontologies with a large central ontology. PROMPT

[21] supports ontology merging, guides users through the process and suggests which classes and properties can be merged and FCA-Merge [22] supports a method for comparing ontologies that have a set of shared instances. IF-Map[23] identifies the mappings automatically by the information flow and generates a logic isomorphism [24].

Based on medical ontologies e.g. UMLS [25], GALEN[5] and GO[6] Health-e-Child is investigating the mapping heuristics for integrated case data. It is evaluating the relative quality of several of these mapping discovery methods for integrated case data. As a consequence it is in the process of providing an optimal combination of the best methods with respect to the accuracy and computation times.

### 4.2 External knowledge

An interesting research area within the scope of the HeC project is the exposure of the HeC knowledge base to the outside world. One example of where this can be useful is to aid in communication between outside sources allowing querying of the HeC system with a user that speaks an outside ontology language. For example, if a common concept is found between the HeC ontology and an external ontology such as the FMA (Foundation Model of Anatomy) [7] then a query, which was originally formulated for the FMA ontology, can also be processed within the HeC domain via the common links that are found between the two ontologies. Much previous work has already been conducted on constructing (semi) automatic mappings between ontologies this being referred to as ontology alignment. The HeC project is currently investigating existing methods and creating new approaches to facilitate the alignment of the HeC ontology with external ontologies. This will facilitate the knowledge sharing of the HeC domain via the ontology and will aid in its reuse.

### 4.3 Semantics support for DSS

Another interesting feature of ontologies is that they can aid in the creation of similarity metrics. This has already been attempted by many projects for example in [26] [27] to gauge the similarity between genes using the GO ontology and by Resnik in [28] to gauge the similarity between different words with the WordNet Thesaurus. This technique can aid in the integration of the other sub-projects in HeC such as the decision support system by creating a similarity metric based on the HeC ontology, hence creating a common base for the training and classification phases of the DSS. One other area where the HeC ontology can be used within the project is the creation of ontology based training data, for example to classify different diseases, this is can be done by using the rule base of the ontology within an expert system [29]. The HeC ontology can be used to annotate different data sets such as images for easy access later, hence creating a semantic image database.

### 4.4　Semantic query enhancement and optimization

Ontologies as mentioned previously can aid in the area of query enhancement. An example is when an image is annotated according to the HeC ontology with the concept of a 'Jaw'. The clinician inputs a simple query into the system, presented here in natural language, stating "Give me all X-Ray images of Jaws for children with a particular disease in a specific age group" then the system will return all of the X-Rays in the database that have been annotated with the concept Jaw. However, if the system uses the power of the HeC ontology it will know that teeth have a 'part-of' relation to a Jaw'. Hence, the system will not only return a result set of images annotated with the concept of Jaw but it will be able to return images annotated with the concept of teeth as well. Therefore, the clinician will be able to take advantage of an enhanced search such as this feature to aid in their experiments.

Query enhancement as the previous example demonstrates is important because it allows the system to provide clinicians with more targeted information. During the requirements analysis phase of the project it became clear that clinicians often struggle to create the complex queries necessary to capture all the data that they require in a study. This may cause too many or too few results to be returned thus undermining the research being undertaken. By using the conceptual model that the HeC ontology provides we can take basic queries from users and translate them into more complex context aware searches. This reduces the amount of time taken by clinicians to locate and group the dataset they require which, in turn minimises the load on the system as fewer searches are necessary. Query optimisation also assists in this regard by using the HeC ontology to aid the creation of efficient data access paths by semantically altering the initial query to find a more efficient execution path within the database. Both query enhancement and optimisation are crucial in delivery of intuitive data access for clinicians whilst at the same time ensuring the scalability and overall stability of the system.

## 5.　Conclusions

In this paper we have used JIA as one disease to illustrate concretely the kind of medical problems we are trying to solve in the HeC project. Many of those are not JIA specific but appear in other areas of medicine possibly with different weights of relevance, importance etc. The Health-e-Child project aims to provide generic solutions without focusing on one particular study. We have selected three considerably different disease areas (paediatric heart diseases, Juvenile Idiopathic Arthritis (JIA) and child brain tumours) to investigate the problems related to differences and commonalities across the paediatric domain. Clinical requirements have been collected during the elicitation sessions with the medical experts and these requirements have been driving the development of the technological solutions to tackle these problems. The integration of the diverse medical data (clinical, epidemiological, imaging, genomic, proteomic etc.) that represent the patient's information at different levels of granularity is very important as the clinical knowledge will span across different medical disciplines allowing clinicians to discover interesting findings and infer new medical knowledge. In addition, the clinical work flows can be quite different in different medical areas (as is exemplified by the three different diseases in our project), but the patient journey can be viewed as the composition of similar tasks (e.g. baseline, diagnosis, treatment, follow-up etc.) for which a common model based on the reusable formalized process patterns should be used. As indicated, future work in the project will enable appropriate knowledge representations including ontologies to be implemented to aid the process of vertical data integration and address the differences across these three disease domains.